# Redundancy Coefficient Gradual Up-weighting-based Mutual Information Feature Selection Technique for Crypto-ransomware Early Detection

**Bander Ali Saleh Al-rimy[1], Mohd Aizaini Maarof[1], Syed Zainudeen Mohd Shaid[1]**

**[1] Faculty of Computing, Universiti Teknologi Malaysia,
81310 UTM Johor Bahru, Johor, Malaysia
bnder321@gmail.com, aizaini@utm.my, szainudeen@utm.my**

## Abstract

Crypto-ransomware is characterized by its irreversible effect even after the detection and removal. As such, the early detection is crucial to protect user data and files of being held to ransom. Several solutions have proposed utilizing the data extracted during the initial phases of the attacks before the encryption takes place. However, the lack of enough data at the early phases of the attack along with high dimensional features space renders the model prone to overfitting which decreases its detection accuracy. To this end, this paper proposed a novel redundancy coefficient gradual up-weighting approach that was incorporated to the calculation of redundancy term of mutual information to improve the feature selection process and enhance the accuracy of the detection model. Several machine learning classifiers were used to evaluate the detection performance of the proposed techniques. The experimental results show that the accuracy of proposed techniques achieved higher detection accuracy. Those results demonstrate the efficacy of the proposed techniques for the early detection tasks.

## I.      Introduction

Although the rapid technological advancements have facilitated the daily activities that people, and organizations conduct, it brought many difficulties and risks as well. Malicious software; known as malware; is one of those threats that obstructs the integration of such technologies into our daily life. Since its occurrence on early 1970s, several types of malware have been witnessed in the wild such as Viruses, Worms, Trojans, Spyware and Ransomware. Ransomware is a malware category that targets user related files and resources and locks them to illicit ransom from the victims if they want their resources back (Azmoodeh et al., 2017; Yalew et al., 2017; Yaqoob et al., 2017; Al-rimy et al., 2018a; Al-rimy et al., 2018b; Chen et al., 2018).

In addition to individual victims, the business entities and governmental institutions are also targeted by ransomware attacks (Al-rimy *et al.*, 2018c; Cohen and Nissim, 2018). It is reported that around $3 million was paid by ransomware victims in 2014 (Homayoun *et al.*, 2017). In 2015, ransomware attackers earned about $352 million around the world (Cohen and Nissim, 2018). Furthermore, Indiana country spent up to $220K in 2016 to recover from ransomware attacks (Cohen and Nissim, 2018). Inability to access data is not the only consequence that ransomware victims incur, the damage could also include downtime costs, loss of money and reputation (Azmoodeh *et al.*, 2017; Al-rimy *et al.*, 2018c).

Two types of ransomware could be distinguished, namely Locking-ransomware and Crypto-ransomware (Al-rimy *et al.*, 2018c; Cohen and Nissim, 2018; Gómez-Hernández *et al.*, 2018). While the former locks user's devices and/or disables some key services in the victim's machine, the latter leverages the cryptography to lock user data and files and hold them to ransom (Chen *et al.*, 2018; Gonzalez and Hayajneh, 2018). The enabling technologies such as of Ransomware-as-a-Service (RaaS), cryptography and difficult to trace Cybercurrency encouraged even non-skilled attackers to develop and disseminate their own crypto-ransomware (Cohen and Nissim, 2018; Gómez-Hernández *et al.*, 2018; Subedi *et al.*, 2018). Consequently, the rate of crypto-ransomware attacks has increased dramatically in recent years (Kharraz et al., 2015; Everett, 2016; Kharraz et al., 2016; Cohen and Nissim, 2018). Contrary to locking-ransomware, the effect of crypto-ransomware attack is irreversible (Gómez-Hernández et al., 2018). That is, if the targeted file was encrypted, the owner would not be able to access it again without the decryption key even after detecting and removing the causing crypto-ransomware (Homayoun et al., 2017; Al-rimy et al., 2018b; Al-rimy et al., 2018c). As such, it is imperative to detect crypto-ransomware attack early, i.e. before the encryption takes place (Homayoun et al., 2017; Yaqoob et al., 2017; Al-rimy et al., 2018a; Al-rimy et al., 2018b; Gómez-Hernández et al., 2018; Rhode et al., 2018).

To detect crypto-ransomware early, several solutions have been proposed which can be categorized as data-centric and process-centric. In data-centric solutions, the focus is on the user data and files subject to attack. Those solutions observe the changes in the file structure and determine whether such changes are suspicious. This approach utilizes several techniques like file entropy, contents similarity measures and decoy techniques (Kharraz et al., 2016; Mbol et al., 2016; Shahriari, 2016; Song et al., 2016; Gómez-Hernández et al., 2018). However, this approach is unable to distinguish whether the change in the file structure was due to crypto-ransomware attack or another benign program such as file compression and/or legitimate cryptography applications (Scaife et al., 2016). Therefore, this approach generates high false alarms rate. More importantly, data-centric approach sacrifices part of user data before detection. This data could be more valuable to victim than the remaining data (Scaife et al., 2016).

On the other hand, resource-centric solutions monitor system resources such as CPU, network, I/O buffer and memory and raises the alarm in case one or more events related to ransomware and/or cryptography were detected. In their study, Cabaj et al. (2015) proposed observing the network traffic to detect the infection chain of Cryptowall. Similarly, the solutions proposed by Cabaj et al. (2017); Cusack et al. (2018) depend on monitoring the network traffic between the victim's machine and ransomware's command and control (C&C) server. Furthermore, Kharraz et al. (2016) put forward UNVEIL system which observes the access patterns in I/O buffer. Similarly, to detect ransomware's suspicious activities, the solution proposed by Song *et al.* (2016) depends on monitoring victim's CPU, I/O and memory. However, relying on ad-hoc events for detection generates high rate of false alarms as those events can be raised by benign programs as well. Moreover, there is no guarantee that those events always precede the encryption due to the variation in the attack strategies of different crypto-ransomware instances (Kharraz *et al.*, 2016).

The idea of building machine learning-based detection models using the early data extracted during the onset of crypto-ransomware attacks was introduced by Sgandurra *et al.* (2016). To define the amount of data required, authors proposed fixed time-based thresholding by which the data captured during the first 30 seconds of ransomware instance runtime were collected and used to build an early detection model. Similarly, Homayoun *et al.* (2017) and Rhode *et al.* (2018) decreased the threshold into 10 seconds and 1 second respectively. Authors of current paper introduced the idea of dynamic thresholding that defines the pre-encryption phase of crypto-ransomware lifecycle (Al-rimy *et al.*, 2018b). Contrary to the fixed thresholding, the proposed technique tracks the pre-encryption phase for each instance individually based on the occurrence of any cryptography-related API. That is, instead of using time-based threshold, this approach uses a threshold based on the cryptography related API calls. However, the small amount of data captured during the initial phases of the attack is one of the challenges that early detection solutions face which causes a poor detection accuracy (Rhode *et al.*, 2018). This problem becomes more complicated with high dimensional feature space as the model becomes prone to overfitting (Reineking, 2016; Fallahpour *et al.*, 2017; Li *et al.*, 2017). One solution to this problem is by conducting features selection to choose the features that represent the underlying dataset and prevents the overfitting (Fallahpour *et al.*, 2017; Li *et al.*, 2017). Several features selection approaches could be used including similarity-based, statistical-based, sparse-learning-based and information theory-based techniques (Li *et al.*, 2017). Characterized by having no assumption about the distribution of data, information theory-based features selection techniques have been utilized by several malware and ransomware detection solutions as well as many other selection problems (Liu *et al.*, 2009; Sgandurra *et al.*, 2016; Wang *et al.*, 2017; Ye *et al.*, 2017).

The theoretical-based features selection techniques like MIFS, mRMR and JMI try to enhance a tradeoff between the relevancy and redundancy terms by adjusting some redundancy coefficients (Brown *et al.*, 2012; Li *et al.*, 2017). Those coefficients were adjusted either to a fixed value or inversely proportional to the size of the selected features set (Battiti, 1994; Yang and Moody, 1999; Hanchuan *et al.*, 2005; Brown *et al.*, 2012; Che *et al.*, 2017). Nevertheless, selecting a fixed value for those parameters is difficult and need to be set experimentally (Brown *et al.*, 2012; Che *et al.*, 2017). On the other hand, the dynamic adjustment of these coefficients changes the belief in the redundancy term at each iteration inversely proportional to the current size of the selected features set (Brown *et al.*, 2012). However, the chance for the candidate feature to be redundant with the features that have been already selected becomes more likely when the size of the selected set increases. As such, the inverse proportional approach that existing techniques employ becomes not suitable and could produce suboptimal set of features which affects the accuracy of the detection model. To this end, this paper proposed the redundancy coefficient gradual up-weighting technique that addresses this issue and calculates the value of redundancy coefficient more accurately. Contrary to existing techniques, the proposed technique increases the value of coefficient proportional to the size of the selected set. Consequently, the belief in redundancy term increases when more features are added to the selected set which improved the relevancy-redundancy tradeoff approximation and produced more distinctive features set. The contribution of this paper is three-folds.

1) A novel redundancy coefficient gradual up-weighting technique was incorporated to the redundancy term of the mutual information to improve the calculation of relevancy-redundancy tradeoff which in turn helpes selecting more informative features set.

2) We have shown that the redundancy term plays a major role into the accuracy of the selected features and the incorporation between the maximum of minimum approach with the proposed redundancy coefficient achieved detection accuracy better than the involvement of conditional redundancy into the calculation.

3) An extensive experimental evaluation was conducted to show the efficacy and significance of the improvement that proposed techniques were contributed to.

For the purpose of this study, crypto-ransomware and ransomware were used interchangeably unless stated otherwise. The rest of this paper is organized as follows. Section II details the methodology and techniques adopted by this study. In Section III, the experimental results were elaborated. Those results were analyzed and discussed in Section IV along with comparison with the related works. The paper was concluded with Section V by a summary of the methods and results as well as suggestions for future work.

## II.     Methodology

In this section, the methodology employed to conduct this study was detailed. The design and implementation of the proposed redundancy coefficient gradual up-weighting technique was presented. Furthermore, the integration of the proposed technique into the mutual information for features selection tasks using the cumulative sum and maximum of minimum was elaborated.

## 1.     Enhance Mutual Information features selection technique.

In this section, the enhanced Mutual Information feature Selection (EMIFS) method is introduced. The proposed method addresses the overestimation issue that conventional MI-based features selection techniques suffer. More particularly, EMIFS controls the degree of belief in the redundancy term based on the size of the selected features set. As such, the degree of belief changes when new features are added. Unlike conventional MIFS, the proposed technique starts with a very low belief in the redundancy term then such belief increases gradually with new features are added. The intuition is that with more features selected, the chance that the new feature is redundant with one or more features in the selected set becomes higher.

## 1.1    Preliminaries

For two discrete variables, the mutual information MI criterion is the amount of information these variables share about each other (Che *et al.*, 2017). This criterion is calculated according to (1) as follows.

$$MI(X;Y) = \sum_{y \in Y} \sum_{x \in X} p(x,y) \log \frac{p(x,y)}{p(x)p(y)} \tag{1}$$

where $p(x)$ and $p(y)$ are the marginal distribution of $x$ and $y$; and $p(x,y)$ is the joint distribution of $x$ and $y$. In their study, Brown *et al.* (2012) proposed a unifying framework for information theoretic features selection by which, several features selections techniques were proposed including Mutual Information Features Selection (MIFS), Information Gain (IG), minimum Redundancy Maximum relevance (mRMR) and Joint Mutual Information (JMI).

According to Brown *et al.* (2012); Li *et al.* (2017), equation (2) represents the general formula of the framework which is referred to as a criterion by linear combinations of Shannon information terms.

$$J(X_k) = MI(X_k;Y) - \beta \sum_{X_j \in S} MI(X_j;X_k) + \gamma \sum_{X_j \in S} MI(X_j;X_k|Y) \tag{2}$$

where $I(X_k;Y)$ is the mutual information between the candidate feature $X_k$ and the class label $Y$; $I(X_j;X_k|Y)$ is the conditional mutual information between the candidate feature $X_k$ and the feature $X_j$ in the selected set $S$ given the class label $Y$; $\beta$ and $\gamma$ are parameters with values between 0 and 1. It turned out that (2) consists of two parts, namely relevancy term represented by (3) and redundancy term represented by (4). Furthermore, the redundancy term consists of two sub-terms, namely the marginal redundancy represented by (5) and the conditional redundancy represented by (6).

$$I(X_k;Y) \tag{3}$$

$$\beta \sum_{X_j \in S} I(X_j;X_k) + \gamma \sum_{X_j \in S} I(X_j;X_k|Y) \tag{4}$$

$$\beta \sum_{X_j \in S} I(X_j;X_k) \tag{5}$$

$$\gamma \sum_{X_j \epsilon S} I\big(X_j; X_k | Y\big) \hspace{6cm} (6)$$

Except IG, all other information theoretical-based features selection techniques try to maximize the tradeoff between the relevancy and redundancy terms. In this context, two types of techniques could be distinguished based on whether or not they include the conditional redundancy term. It turned out that the calculation of relevancy term is same in all techniques and involves calculating the relevancy between the candidate feature $X_k$ and the class label $Y$. As such, the difference in the performance between those techniques comes from the redundancy calculation. That is, the feature's relevancy was determined by relevance term and to some extend by the conditional redundancy as well.

As shown in (2), the values of $\beta$ and $\gamma$ play an important role in the relevancy-redundancy tradeoff which determines the feature's significance. Concretely, the small value of $\beta$ contributes to decreasing the effect of the redundancy which; consequently; increases the feature's significance whereas the small value of $\gamma$ decreases such significance. As mentioned previously, existing techniques either assign fixed values for $\beta$ and $\gamma$ or dynamic values based on the size of the selected feature set at each iteration. However, the selection of those parameters is difficult and need to be set experimentally (Brown *et al.*, 2012; Che *et al.*, 2017). On the other hand, some techniques such as mRMR and JMI assign the values of these parameters dynamically based on the current size of the selected features set (Yang and Moody, 1999; Hanchuan *et al.*, 2005). Such approach changes the belief in the redundancy term at each iteration inversely proportional to the current size of the selected features $S$ (Brown *et al.*, 2012).

## 1.2    Redundancy Coefficient Gradual Up-weighting Technique

To evaluate the suitability of the feature to represent the underlying dataset, the mutual information criterion was used. The higher MI value the more suitable the feature is. Such criterion is calculated according to (7) as follows.

$$J(x_k) = MI(x_k, y) - \beta \sum_{s_j \in S} I(x_k, x_j) \qquad (7)$$

where $x_k$ denotes the candidate feature; $s_j \in S = \{s_1, s_2, \ldots, s_m\}$ is the $j^{th}$ feature in the set of already selected features $S$; and $\beta$ is a nonnegative parameter between 0 and 1. The left term in the equation represents feature relevancy whereas the right term represents feature redundancy. The value of $\beta$ determines the strength of belief in the redundancy term and is calculated according to (8) as follows.

$$\beta = \frac{|S|}{|F|} \qquad (8)$$

where $|S|$ and $|F|$ denote the number of features in the selected and original set respectively. Therefore, EMIFS selects the informative features according to (9).

$$J(x_k) = MI(x_k, y) - \frac{|S|}{|F|} \sum_{s_j \in S} I(x_k, x_j) \qquad (9)$$

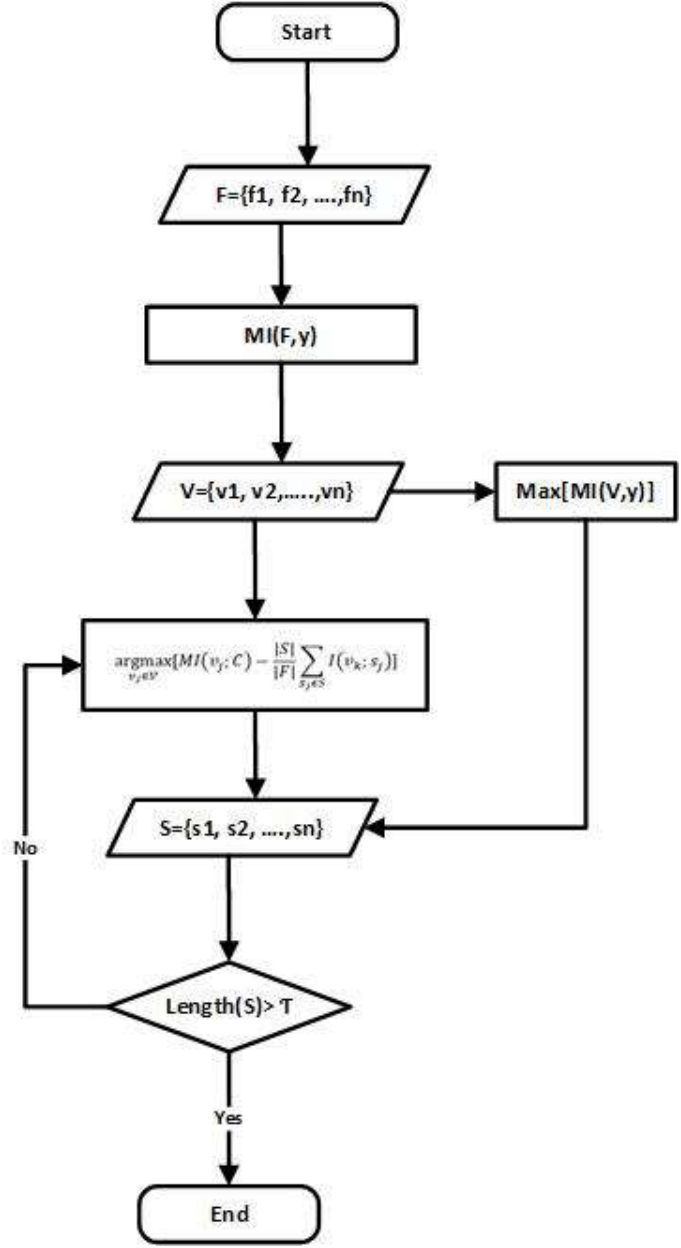

**Figure 1:** The EMIFS Technique

Given the feature vector $F$ built in the previous section, EMIFS selected the informative features according to (9). In the beginning of selection process, the mutual information value (MI) for each feature in $F$ was calculated with respect to the class label. After that, the feature with the highest MI value was chosen and stored in the selected set $S$. Then, the next and subsequent features were chosen according to (9). Then, the features in $S$ were ranked from high to low based on $J(x_k)$ value. Finally, the top $\tau$ features were retained, and the others were removed from $S$. The value of $\tau$ is specified based on the desired number of features. Figure 1 shows the design of EMIFS while Figure 2 illustrates its pseudo code.

As shown in Figure 2, $F = \{f_1, f_2, f_3, \dots f_{n-1}, f_n\}$ is the original features vector with $n$ number of features; $V$ is temporary set that holds the features whose MI value has been already calculated; $S = \{s_1, s_2, \dots, s_\tau\}$ is the selected set with $\tau$ number of features. EMIFS started by initializing empty $V; S$ sets and calculating the MI value for each feature $f_i$ in $F$. Based on the MI value, those features were ranked and stored in the set $V$. Then, the feature $v_k$ in $V$ with $\max(V, MI)$ was removed from $V$ and added into $S$. The next feature $v_p$ was chosen according to (10).

$$v_p = \underset{v_j \in V}{\mathrm{argmax}}[MI(v_j; C) - \frac{|S|}{|F|}\sum_{s_j \in S} I(v_k; s_j)] \tag{10}$$

Equation (12) shows that at each iteration, the feature $v_j$ from $V$ that produces the highest mutual information with the class label given the already selected features was added to $S$. When $\tau$ was reached, the selection process stopped.

**EMIFS Method**

Input: $F = \{f_1, f_2, \dots, f_n\}$ original features vector; $C$ is the class label, $\tau$ is the number of required features.

Output: $S = \{s_1, s_2, \dots, s_p\}$ the final features set.

1. $V \leftarrow \emptyset; S \leftarrow \emptyset$
2. for each feature $f_i \in F$:

      $v_i = MI(f_i; C)$

      $V \leftarrow V \cup v_i$

3. $v_k \leftarrow \max(V, MI)$
4. $S \leftarrow v_k; V \leftarrow V \backslash \{v_k\}$
5. for $\forall (v_j, s_m)$ with $v_j \in V$ and $s_m \in S$
6.     compute $MI(C; s_m | v_j)$
7.     $v_p = \underset{v_j \in V}{\mathrm{argmax}}[MI(v_j; C) - \frac{|S|}{|F|}\sum_{s_j \in S} I(v_k; s_j)]$
8.     $V \leftarrow V \backslash \{s_p\}$
9.     $S \leftarrow S \cup \{s_p\}$
10. Repeat $6 - 9$ while $length(S) \leq \tau$

**Figure 2:** The Pseudo Code of EMIFS Technique

## 2    Maximum of Minimum-Based Enhanced Mutual Information Features Selection (MM-EMIFS) Technique

In this section an improvement to EMIFS is introduced by employing the maximum of minimum approximation approach for features' significance estimation. This approach extends the calculation of $MI(x_i, s_j)$ to $MI(x_i, S)$. As mention by Che *et al.* (2017), the overestimation of redundancy term weakens the relevancy term while the underestimation does not have effect. As such, MM-EMIFS applies the maximum of minimum approximation on the redundancy term so that it alleviates the issue of redundancy overestimation caused by the cumulative sum that is used by the existing solutions. This approximation relaxes the redundancy calculation without affecting the relevancy term. Therefore, MM-EMIFS is able to produce better estimation for features relevance.

In the onset of selection process, the mutual information value for each feature in the original features set was calculated. Then, the feature with higher MI value was stored in the selected set $S$. The next and subsequent features were chosen using the maximum of minimum approximation. At each iteration, the feature with highest mutual information score and minimum redundancy value was added into $S$. Figure 3 illustrates the pseudo code of MM-EMIFS technique.

As shown in Figure 3, $F = \{f_1, f_2, f_3, \dots f_{n-1}, f_n\}$ is the original features vector with $n$ number of features; $V$ is temporary set that holds the features whose MI value has been already calculated; $S = \{s_1, s_2, \dots, s_\tau\}$ is the selected set with $\tau$ number of features. The process starts by finding the feature with highest MI value, adding it to the final features $S$ set and removing it from the original set $F$. Equation (11) shows that, at each iteration a greedy search was conducted so that for each feature $f_m$ in the candidate set $F$ the mutual information between that particular feature and the class label given each feature in the selected set $S$. Then, the minimum MI value is recorded in a list $L_{min}$. At the end of this process, the feature corresponding to the maximum value in $L_{min}$ is selected and added into the final set $s$ and removed from the candidate set $V_{new}$.

$$V_k = \underset{V_j \in V_{new}}{\arg\max} \, \underset{V_i \in S}{\min} \{MI(C; V_j | V_i)\} \qquad (11)$$

The process continued until $S$ became full of the features.

**EMIFS Method**

Input: $F = \{f_1, f_2, \ldots, f_n\}$ original features vector; $C$ is the class label, $\tau$ is the number of required features.

Output: $S = \{s_1, s_2, \ldots, s_p\}$ the final features set.

1. $V \leftarrow \emptyset; S \leftarrow \emptyset$
2. for each feature $f_i \in F$:
   $\quad\quad v_i = MI(f_i; C)$
   $\quad\quad V \leftarrow V \cup v_i$
3. $v_k \leftarrow \max(V, MI)$
4. $S \leftarrow v_k; V \leftarrow V \setminus \{v_k\}$
5. for $\forall (v_j, s_m)$ with $v_j \in V$ and $s_m \in S$
6. $\quad\quad$ compute $MI(C; s_m | v_j)$
7. $\quad\quad s_t = \arg\max_{v_j \in V} \min_{s_r \in S} \{MI(C; v_j | s_r)\}$
8. $\quad\quad V \leftarrow V \setminus \{s_p\}$
9. $\quad\quad S \leftarrow S \cup \{s_p\}$
10. Repeat $6 - 9$ while $length(S) \leq \tau$

**Figure 3:** The Pseudo Code of MM-EMIFS Technique

## III.    Results

In this section, a brief description about the setup of experimental environment in which the implementation of the proposed techniques were conducted was given. Then the dataset used by this study was detailed explaining the different instances and the type of data that have been acquired. The experimental results of each technique were introduced including the comparison with the previous studies as well.

## A.      The Experimental Environment Setup

The experiments were conducted on a controlled environment built on a machine with Intel(R) Core(TM) i7-4790 CPU @ 3.60 GHZ and 16 GB RAM. The analysis environment was built according to Rossow *et al.* (2012). The Cuckoo Sandbox; a well-known and widely used malware analysis platform; was used as analysis environment (Wang and Wang, 2015; Stiborek *et al.*, 2018b; Stiborek *et al.*, 2018a). The VMware Technology was utilized to build the sandbox.  Within this sandbox, the host machine was created using Linux ubuntu 4.4.0-59-generic. Then, the VirtualBox was utilized to create the gust machine using MS Windows 32-bit guest machine. Crypto-ransomware and benign programs were run one by one. For each program, the data were dumped into an independent trace file. Those trace files contain the API calls used by the program under analysis during the runtime. After each run, the gust machine was restored into the original, clean state. Extracted data was gathered and the features were extracted and selected during the preprocessing phase. Once ready, the dataset was used to build the detection model. The proposed techniques as well as results and analysis were implemented using Python libraries including Sklearn, Pandas and Numpy.

## B.      The Dataset

The corpus of crypto-ransomware binaries used in this study were downloaded from virusshare.com public repository (Sgandurra *et al.*, 2016; Chen *et al.*, 2017; Christensen and Beuschau, 2017; Le Guernic and Legay, 2017). The corpus consists of 8,152 samples. These samples represent different families such as Cerber, TeslaCrypt, CryptoWall, Petya and WannaCry. Those samples were collected during the period from Sep 2016 to Aug 2017. In addition, 1000 benign programs were downloaded from informer.com (Pandey and Mehtre, 2015; Sgandurra *et al.*, 2016; Chen *et al.*, 2017; Ioanid *et al.*, 2017), a popular Windows-based applications repository. Then, both ransomware and benign programs were run in the sandbox. After submitting the sample to the analyzing machine, the sandbox agent in gust machine hooks the process created by that sample and captures the APIs along with the parameters and dumps them into a trace file in the host machine specified for that sample. These files constitute the corpus by which the dataset was built and the features were extracted and selected. From these files, the pre-encryption dataset was built according to Al-rimy *et al.* (2018b)

## C.     Experimental Results of EMIFS

To evaluate the performance of the proposed EMIFS, the pre-encryption dataset underwent features extraction using the TF-IDF technique which produced the pre-encryption features vector. After that, EMIFS was used to select the most informative features for the pre-encryption phase of crypto-ransomware lifecycle. The experiments were conducted using several feature sets with different number of features, i.e. 5, 10, 15, 20, 25, 30, 35, 40, 45, and 50 features. The dataset was divided into training set and testing set using 10-fold cross validation approach. Several machine learning algorithms were used in this evaluation including Support Vector Machine (SVM), Logistic Regression (LR), Decision Tree (DT), K-Nearest Neighbor (KNN), Random Forest (RF), adaBoost, and Multi-layer Perceptron (MLP). The testing set then was used to determine the classification performance of those classifiers. The accuracy metric was used to measure the performance of EMIFS on those algorithms. This accuracy was calculated according to (12).

$$Accuracy = \frac{tp + tn}{tp + tn + fp + fn} \qquad (12)$$

where $tp, tn, fp, fn$ denote the true positive, true negative, false positive and false negative respectively.

Table 1 shows the accuracy results of each classifier. Each raw in the table corresponds to one features set used to train different classifiers. It can be observed that for all classifiers, the classification accuracy increases with the increase in the size of features set especially the sets with less than 30 features. When the size of features set goes beyond 30, the increase becomes less gradual and sometimes the classifiers experience a slight accuracy drop. Furthermore, the accuracy average (Avg.) of all features sets per classifier ranges between 0.9183 and 0.9708.

**Table 1:** Experimental results of the EMIF on pre-encryption dataset (extracted according to Al-rimy *et al.* (2018b)) with different sizes of features sets used to train several classifiers

|      | LR     | SVM    | DT     | RF     | KNN    | adaBoost | MLP    |
|------|--------|--------|--------|--------|--------|----------|--------|
| 5    | 0.9039 | 0.9286 | 0.9643 | 0.9714 | 0.9511 | 0.9386   | 0.9107 |
| 10   | 0.9168 | 0.9493 | 0.9661 | 0.9711 | 0.9518 | 0.9432   | 0.9207 |
| 15   | 0.9203 | 0.9518 | 0.9654 | 0.9707 | 0.9518 | 0.9432   | 0.9239 |
| 20   | 0.9203 | 0.9514 | 0.9664 | 0.9682 | 0.9518 | 0.9432   | 0.9239 |
| 25   | 0.9203 | 0.9514 | 0.9654 | 0.9722 | 0.9522 | 0.9432   | 0.9221 |
| 30   | 0.9203 | 0.9514 | 0.9661 | 0.9707 | 0.9522 | 0.9432   | 0.9207 |
| 35   | 0.9203 | 0.9514 | 0.9675 | 0.9725 | 0.9518 | 0.9432   | 0.9228 |
| 40   | 0.92   | 0.9514 | 0.9679 | 0.9707 | 0.9518 | 0.9432   | 0.9207 |
| 45   | 0.92   | 0.9514 | 0.9654 | 0.9718 | 0.9511 | 0.9432   | 0.9211 |
| 50   | 0.9203 | 0.9507 | 0.965  | 0.969  | 0.9515 | 0.9389   | 0.9243 |
| **Avg.** | 0.9183 | 0.9489 | 0.9659 | 0.9708 | 0.9517 | 0.9423 | 0.9211 |

## D.    Experimental Results of MM-EMIFS

To evaluate the prediction accuracy of the proposed MM-EMIFS technique, an experimental evaluation was conducted by applying the proposed technique on the pre-encryption dataset. The experiments were carried out using several machine learning classifiers such as LR, SVM, DT, RF, KNN, adaBoost and MLP. Furthermore, different features set sizes were used ranging from 5 to 50 features. In addition, the dataset was divided into training and testing sets using 10-fold cross-validation approach. Detection accuracy was used to measure the performance of the proposed technique according to (12).

Table 2 shows the experimental of from applying MM-EMIFS technique on the pre-encryption dataset. It can be observed that the classification accuracy increases proportional to the number of features. It can also be noticed that the increase in accuracy decelerates starting from the features set size 30 and above. Furthermore, the accuracy average (Avg.) of all features sets per classifier ranges between 0.9385 for and 0.98015.

**Table 2:** Experimental results of the MM-EMIF on pre-encryption dataset ( extracted according to Al-rimy *et al.* (2018b)) with different sizes of features sets used to train several classifiers

|      | LR     | SVM    | DT     | RF      | KNN    | adaBoost | MLP    |
|------|--------|--------|--------|---------|--------|----------|--------|
| 5    | 0.9039 | 0.9286 | 0.955  | 0.9647  | 0.94   | 0.9296   | 0.9064 |
| 10   | 0.9111 | 0.9536 | 0.9607 | 0.9654  | 0.9489 | 0.9454   | 0.9304 |
| 15   | 0.9221 | 0.965  | 0.9743 | 0.9829  | 0.9625 | 0.9711   | 0.9436 |
| 20   | 0.9329 | 0.9639 | 0.9732 | 0.9843  | 0.9618 | 0.9739   | 0.9482 |
| 25   | 0.9439 | 0.9722 | 0.9729 | 0.9832  | 0.9657 | 0.9725   | 0.9564 |
| 30   | 0.9529 | 0.9729 | 0.9757 | 0.9843  | 0.9664 | 0.9747   | 0.9618 |
| 35   | 0.9543 | 0.9714 | 0.9764 | 0.9843  | 0.9657 | 0.975    | 0.9611 |
| 40   | 0.9546 | 0.9697 | 0.9764 | 0.9832  | 0.9661 | 0.9754   | 0.9593 |
| 45   | 0.9539 | 0.9693 | 0.9739 | 0.9832  | 0.9661 | 0.9754   | 0.9607 |
| 50   | 0.9557 | 0.9693 | 0.9761 | 0.9861  | 0.9661 | 0.9754   | 0.9607 |
| **Avg.** | 0.9385 | 0.9636 | 0.9715 | 0.98015 | 0.9609 | 0.9668   | 0.9489 |

### E.    Comparison with the Related Techniques

To show the efficacy of the proposed EMIFS technique, the results were compared with three of information theoretical-based features selection techniques, namely Mutual Information Features Selection (MIFS) and Minimum Redundancy Maximum Relevance (mRMR) and Joint Mutual Information Maximization (JMIM). Several machine learning classifiers were used in this evaluation, i.e. LR, SVM, DT, RF, KNN, adaBoost and MLP. Moreover, the experiments were conducted using different sizes of features sets ranging from 5 to 50 and incremented by 5 features between each two consequent sets. The classification accuracy was used as the measurement of the classification performance. The Figure 4 shows the comparison results between EMIFS and related techniques. Based on the comparison results, both proposed EMIFS and MM-EMIFS outperforms MIFS and mRMR. Moreover, the accuracy of MM-EMIFS is higher than that of the other techniques. It is worth noticing that the accuracy of all techniques increases with the size of the features set until the size approaches 30 features then starts to stabilize and the increase in the accuracy becomes less gradual.

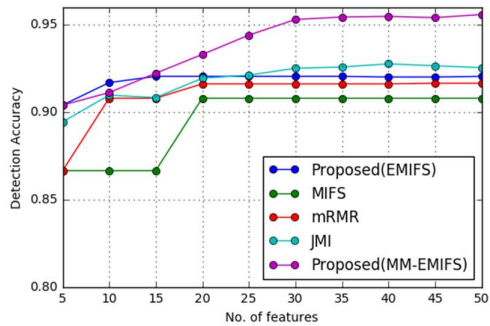

(LR)

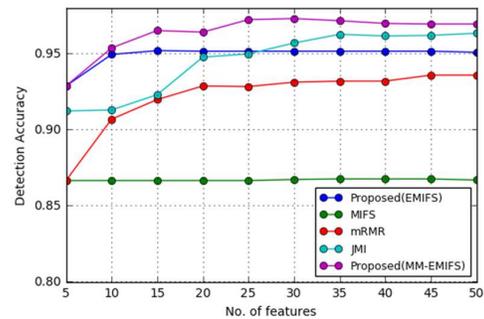

(SVM)

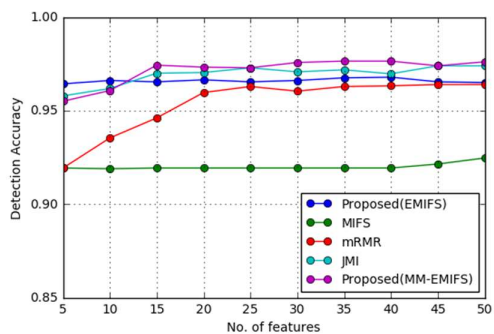

(DT)

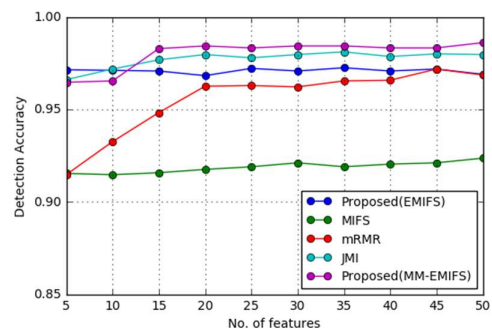

(RF)

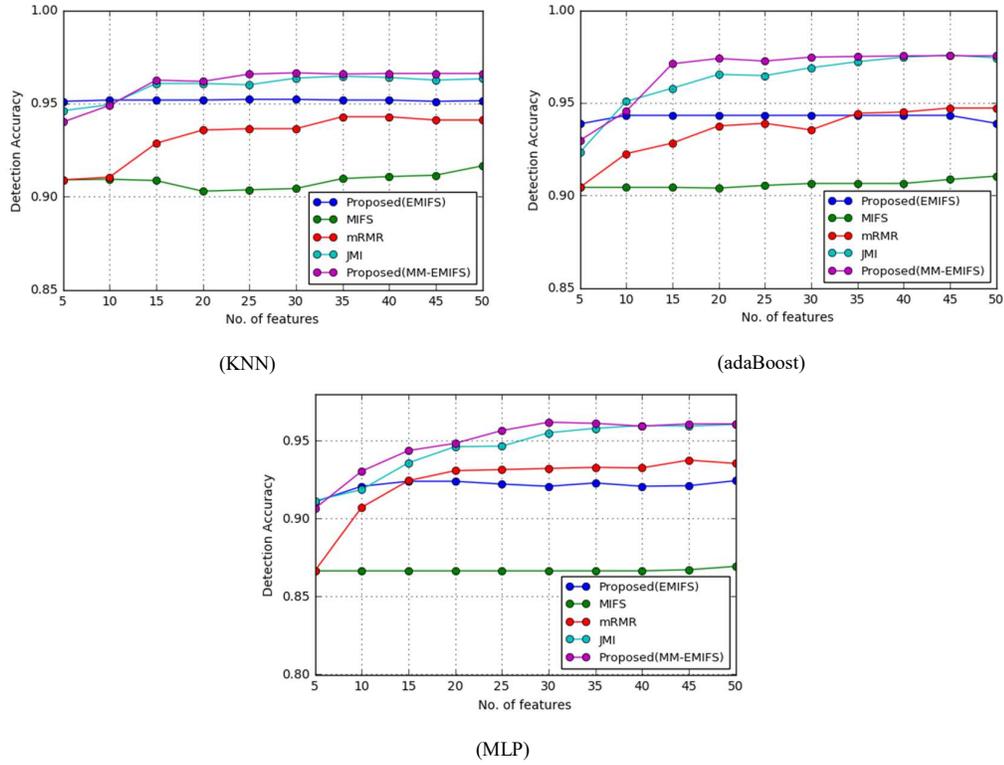

(KNN)

(adaBoost)

(MLP)

**Figure 4:** The comparison between proposed EMIFS, MM-EMIFS and the related techniques using the pre-encryption dataset (DR-pre)

To find out how significant the improvement is, t-test was carried out on the classification accuracy results of the each proposed technique and the related techniques. The significance level was adjusted to the standard value ($\alpha = 0.05$). Table 3 shows the results of the significance test between EMIFS and the related techniques. It shows that the improvement is statistically significant all algorithms except the adaBoost with mRMR. It also suggests that despite the superiority that mRMR was shown when used by MLP algorithm, the accuracy difference with the proposed EMIFS on the same algorithm was not significant. In addition, Table 4 shows that that the improvement of the proposed MM-EMIFS was significant in all cases.

**Table 3**: Significance test (t-test) results between the proposed EMIFS technique and the related techniques with different classification algorithms

| | MIFS | | | mRMR | | |
|---|---|---|---|---|---|---|
| | *t-value* | *p-value* | *Significant?* | *t-value* | *p-value* | *Significant?* |
| **LR** | 4.172839 | 0.001201 | Yes | 2.636869 | 0.013526 | Yes |
| **SVM** | 36.90325 | 1.95E-11 | Yes | 5.865621 | 0.00012 | Yes |
| **DT** | 60.47135 | 2.33E-13 | Yes | 2.590326 | 0.014599 | Yes |
| **RF** | 48.34745 | 1.74E-12 | Yes | 2.616883 | 0.013977 | Yes |
| **KNN** | 31.37098 | 8.34E-11 | Yes | 4.812531 | 0.000478 | Yes |
| **adaBoost** | 35.84226 | 2.53E-11 | Yes | 1.77984 | 0.054402 | No |
| **MLP** | 45.53682 | 2.97E-12 | Yes | -0.33514 | 0.372599 | No* |

**Table 4**: Significance test (t-test) results between the proposed MM-EMIFS technique and the related techniques with different classification algorithms

| | MIFS | | | mRMR | | |
|---|---|---|---|---|---|---|
| | *t-value* | *p-value* | *Significant?* | *t-value* | *p-value* | *Significant?* |
| **LR** | 16.30105 | 2.73E-08 | Yes | 7.061605 | 2.96E-05 | Yes |
| **SVM** | 22.97778 | 1.33E-09 | Yes | 15.61557 | 3.98E-08 | Yes |
| **DT** | 22.8948 | 1.37E-09 | Yes | 6.335615 | 6.76E-05 | Yes |
| **RF** | 6.904936 | 3.51E-05 | Yes | 6.904936 | 3.51E-05 | Yes |
| **KNN** | 16.68077 | 2.24E-08 | Yes | 17.85193 | 1.23E-08 | Yes |
| **adaBoost** | 12.76693 | 2.27E-07 | Yes | 16.04857 | 3.13E-08 | Yes |
| **MLP** | 14.5606 | 7.3E-08 | Yes | 13.1389 | 1.77E-07 | Yes |

## IV.    Discussion

In this study, the redundancy coefficient gradual up-weighting technique for mutual information-based features selection was introduced. Furthermore, the proposed technique was integrated into both MI-based features selection types, namely the cumulative sum-based techniques and the maximum of minimum-based techniques. In this section, the results of experimental evaluation of the proposed EMIFS and MM-EMIFS techniques are discussed. The discussion focuses on the improvement of the classification accuracy these techniques have achieved as well as the significance of such improvement. The section starts discussing EMFIS technique then proceeds to MM-EMFIS.

## A. The EMIFS Technique

In this study, the issue of redundancy term overestimation was investigated. To that, two approaches were examined, i.e. gradual redundancy up-weighting and maximum of the minimum. First, the concept of gradual redundancy up-weighting was introduced. The approach starts with minimum weight of redundancy between the candidate feature and the features already selected. This weight increases with more features are added into the selected features set. Moreover, the maximum of minimum approach was employed to mitigate the issue of redundancy term overestimation caused by cumulative sum (CUMSUM) approach currently used by the related techniques. Based on these approaches, EMIFS and MM-EMIFS were proposed. The improvement in the classification accuracy demonstrates the ability of the proposed techniques to select the informative features that represent the behavior of crypto-ransomware even with only limited amount of data at the early phases of the attack. In this section, the classification performance results of the proposed EMIFS and MM-EMIFS techniques are analyzed and discussed in addition to pros and cons in the light of the results shown in Sction III.

The results in Table 1 emphasize the efficacy of the gradual redundancy up-weighting approach in selecting the informative features of the pre-encryption phase. These results suggest that using the gradual redundancy up-weighting, EMIFS becomes able to adjust the redundancy-relevancy tradeoff more effectively. The high level of accuracy that EMIFS starts with indicates that the feature relevancy plays the main rule at the early stages of the selection process. Likewise, it can be noticed that the increase of classification accuracy decelerates when number of selected features exceeds 30. This indicates that the redundant information becomes more influential on the MI calculation when more features are added into the selected features set. This advocates the assumption that the influence of redundant features starts low and increases gradually with more features are added into the selected features set. As such, the proposed technique deals with this situation more effectively.

The comparison results shown in Figure 4 suggest that the proposed gradual redundancy up-weighting approach employed by EMIFS is more effective than the approach used in the existing techniques which decreases the influence of redundancy term with the growth of the selected features set size. Similar to related techniques, the results show that the increase of classification accuracy of EMIFS becomes less gradual when the features set's size becomes large which is the This indicates that the redundancy up-weighting approach does not experience deterioration in the classification accuracy with the growth of the selected features size. It is worth noticing that even though EMIFS does involve for the conditional redundancy term, it shows comparable results with the JMI that count for such term. This indicates that the conditional redundancy has inferior influence on the MI calculation to the marginal redundancy.

## B. MM-EMIFS Technique

The results in Table 2 shows that the integration between the gradual redundancy up-weighting and the maximum of minimum employed by MM-EMIFS produced better classification accuracy. This indicates that, by enhancing the calculation of minimum MI value between the candidate feature and the features in the selected set the gradual redundancy up-weighting improved the maximum of minimum approximation. Such improvement helps mitigating the underestimation that maximum of minimum approach suffers. As such, the feature corresponding to the maximum value of the minimum MIs was more informative than the one selected by the traditional maximum of minimum technique.

The comparison shown in Figure 4 suggests that the integration between gradual redundancy up-weighting and the maximum of minimum approaches produced better classification accuracy than existing techniques. It is worth noticing that the classification accuracy of the proposed MM-EMIFS was better than JMI technique which involves the conditional redundancy term in the MI calculation. This can be seen by comparing the accuracy results of the proposed technique with JMI using several classifiers. This indicates that, with the enhanced calculation of redundancy term using gradual redundancy up-weighting, the need to calculate the conditional redundancy term become of less importance. Consequently, the proposed MM-EMIFS produces better accuracy with less computational complexity.

Tables 3 and 4 show the results of t-test between the existing techniques with EMIFS and MM-EMIFS respectively using several machine learning classifiers. The results were less than the significance level in the vast majority of cases. This indicates that the improvement that proposed gradual redundancy up-weighting has made to the calculation of relevancy-redundancy tradeoff in the MI calculation was significant. One limitation of the proposed EMIFS and MM-EMIFS is that the denominator of $\alpha$ is fixed to the number of features in the input space. More enhancement could be done by considering dynamic value that changes based on the current situation of the selection process. Furthermore, the existing solutions focus the tradeoff between the relevancy and redundancy and overlook the tradeoff between the marginal redundancy and conditional redundancy terms. Therefore, the involvement of conditional redundancy term could enhance the performance further if a suitable tradeoff between $\alpha$ and $\beta$ in the equation () was determined. Such involvement could also open another direction to study the tradeoff between the three elements, i.e. relevancy, marginal redundancy and conditional redundancy.

## V.    Conclusion

In this paper, the redundancy gradual up-weighting Mutual Information-based features selection technique for crypto-ransomware early detection was proposed. The redundancy-relevancy tradeoff in the feature selection process was improved by integrating the redundancy gradual up-weighting approach into the redundancy term of the mutual information features selection. The detection accuracy of the proposed redundancy gradual up-weighting approach was higher than those of the related techniques as well. The results show the efficacy of the proposed techniques for crypto-ransomware early detection. These techniques could be applied for early detection of other attacks such as malware detection and intrusion detection. One limitation of the proposed redundancy coefficient gradual up-weighting technique is the inconsideration of conditional redundancy term when calculating the feature significance. Currently, we are working on applying the similar approach on the conditional redundancy term to further improve the features selection process and enhance the detection accuracy.